# Simulating Maxwell-Schrödinger Equations by High-Order Symplectic FDTD Algorithm

Guoda Xie, Zhixiang Huang, *Senior Member, IEEE*, Ming Fang, and Wei E.I. Sha, *Senior Member, IEEE*

*Abstract*—A novel symplectic algorithm is proposed to solve the Maxwell-Schrödinger (M-S) system for investigating light-matter interaction. Using the fourth-order symplectic integration and fourth-order collocated differences, Maxwell-Schrödinger equations are discretized in temporal and spatial domains, respectively. The symplectic finite-difference time-domain (SFDTD) algorithm is developed for accurate and efficient study of coherent interaction between electromagnetic fields and artificial atoms. Particularly, the Dirichlet boundary condition is adopted for modeling the Rabi oscillation problems under the semi-classical framework. To implement the Dirichlet boundary condition, image theory is introduced, tailored to the high-order collocated differences. For validating the proposed SFDTD algorithm, three-dimensional numerical studies of the population inversion in the Rabi oscillation are presented. Numerical results show that the proposed high-order SFDTD(4,4) algorithm exhibits better numerical performance than the conventional FDTD(2,2) approach at the aspects of accuracy and efficiency for the long-term simulation. The proposed algorithm opens up a promising way towards a high-accurate energy-conservation modeling and simulation of complex dynamics in nanoscale light-matter interaction.

## I. INTRODUCTION

Computational electromagnetics (CEM) [1] is fundamental to the electronic technology and plays an essential role in the field of modeling and designing modern electronic devices. The fast innovation process and increasing complication of technology require higher integration and smaller component dimensions, leading to new modeling and simulation challenges in the traditional CEM technology. For instance, advances in nanotechnology enable to fabricate devices at nanoscale, where quantum mechanics (QM) effects become more significant and cannot be ignored. As electrons are trapped in a deep subwavelength scale, the electronic states will be quantized and thus the dynamics of electrons should be governed by Schrödinger equations [2]. Meanwhile, the excited electrons produce quantum current which generates EM fields. Hence, in order to precisely describe the light-matter interaction between EM fields and artificial atoms, the coupled Maxwell-Schrödinger (M-S) system must be considered. In fact, the coupled M-S system has been used to solve the multi-physics problems in literatures. In [3], [4], a transmission line matrix technique combined with the finite-difference time-domain (FDTD) approach [5], [6] is presented to solve the M-S system, which is applied to describe the EM characteristics of nanostructures. In [7], a M-S numerical toolbox is proposed to study the interaction between electrons wave packet and gratings. In addition, some researchers studied the optical problems and nonlinear effects in plasmon/molecule interaction by solving the coupled M-S equations [8], [9], [10], [11], [12], [13]. Among these methods, the EM fields (**E** and **H**) in Maxwell's equations are included in the numerical system. However, it revealed that numerical simulations based on the **E-H** formulations may suffer from low-frequency breakdown when the incident wavelength is much larger than the object's size [14], [15].

To circumvent this problem, some efforts have been made to solve the coupled M-S equations where only potentials are considered rather than the **E-H** equations [16], [17]. Recently, a coupled M-S system for modeling electromagnetic fields-artificial atoms interaction has been introduced in [18]. The Coulomb gauge is employed to solve the Maxwell equation in the form of vector and scalar potentials. In addition, the reduced eigenmode expansion (REE) approach is used to expand the atomic wave function, which can be connected to the Bloch equation [19], [20]. This coupled system is solved by the conventional FDTD approach. Nevertheless, the standard FDTD approach only has second-order accuracy in both space and time, significant error accumulation will occur in the simulation due to its large numerical dispersion and anisotropy. Therefore, dense grids should be adopted to satisfy the precision requirement, which will result in substantial memory and computational time, especially for the long-term numerical simulation. It should be noted that the M-S system always needs more time steps compared to a pure EM system. Based on the above descriptions, developing a more accurate

This work was supported by the NSFC (Nos. 61722101, 61801002, 61701001, 61701003), Open Fund for Discipline Construction, Institute of Physical Science and Information Technology, Anhui University (No.2019AH001), USFC of Anhui Province (Nos. KJ2018A0015, KJ2017ZD51, KJ2017ZD02),and Marie Skłodowska-Curie Individual Fellow ship (MSCA-IF-EF-ST(752898)) .

Guoda Xie, Zhixiang Huang and Ming Fang are with the Key Laboratory of Intelligent Computing and Signal Processing, Ministry of Education, Anhui University, Hefei 230601, China (e-mail: zxhuang@ahu.edu.cn (ZXH)).

Wei E.I. Sha is with Key Laboratory of Micro-nano Electronic Devices and Smart Systems of Zhejiang Province, College of Information Science and Electronic Engineering, Zhejiang University, Hangzhou, 310027, China. (email: weisha@zju.edu.cn)

energy-preservation solution for the M-S system is highly important to model and optimize emerging nanodevices. Fortunately, intensive research works suggested that symplectic algorithm is one of the best methods for long-term simulation, in view of its energy-conserving and highly stable characteristics [21], [22]. Furthermore, numerous physical systems can be modeled by Hamiltonian differential equations, the time evolution of the differential equations is symplectic transformation [23], [24]. The symplectic algorithm is a differential method, which is based on the basic principles of Hamiltonian mechanics for preserving symplectic structures of Hamiltonian systems.

In this paper, we present a three-dimensional high-order symplectic FDTD algorithm (SFDTD(4,4)), with the fourth-order symplectic propagators to solve the coupled M-S system. The M-S system will be used to describe the Rabi oscillation phenomenon generated by artificial atoms under the influence of external EM fields. Simultaneously, for the fourth-order collocated spatial differences, the image theory [25] is adopted to handle the Dirichlet boundary condition. At last, the calculations of the atomic population inversion have been carried out to demonstrate the superiorities of the high-order SFDTD(4,4) algorithm at the respects of calculation precision and efficiency.

## II. FORMULAE

### A. Symplectic Structure of the M-S System

For QM part, an electron with charge $e$ and mass $m$ interacting with external fields can be described by a modified Schrödinger equation

$$i\hbar \frac{\partial \psi(\mathbf{r}, t)}{\partial t} = \left\{ \frac{1}{2m} [\mathbf{p} - e\mathbf{A}(\mathbf{r}, t)]^2 + e\phi(\mathbf{r}, t) + V(\mathbf{r}) \right\} \psi(\mathbf{r}, t) \quad (1)$$

where $\mathbf{p} = -i\hbar \nabla$ denotes the canonical momentum operator, and the variables $\mathbf{A}(\mathbf{r}, t)$ and $\phi(\mathbf{r}, t)$ are the vector and scalar potentials, respectively. $V(\mathbf{r})$ is the electrostatic confinement potential. Hence, the Hamiltonian operator of an electron bound to an atom under EM fields radiation is given by

$$H = \frac{1}{2m} [\mathbf{p} - e\mathbf{A}(\mathbf{r}, t)]^2 + e\phi(\mathbf{r}, t) + V(\mathbf{r}) \quad (2)$$

Since the vector potentials ($\mathbf{A}$) and scalar potentials ($\phi$) can be directly incorporated into the EM-QM hybrid system, the classical Maxwell equation based on the $\mathbf{E}$-$\mathbf{H}$ components are replaced by the $\mathbf{A}$-$\phi$ components. Additionally, the Coulomb gauge ($\nabla \cdot \mathbf{A}$) which is often used to analyze the quantum optics problem [26] is adopted. After this, an auxiliary variable $\mathbf{Y}$ is defined for the EM system ($\mathbf{Y}=\varepsilon_r\varepsilon_0\mathbf{E}$, $\varepsilon_r$ is the relative permittivity).

According to (1), (2) and the auxiliary variable defined in EM system, the coupled Hamiltonian in equation (2) can be written in the following expression

$$H(\mathbf{A}, \mathbf{Y}, \psi, \psi^*) = H^{em}(\mathbf{A}, \mathbf{Y}) + H^q(\mathbf{A}, \psi, \psi^*) \quad (3)$$

where

$$H^{em}(\mathbf{A}, \mathbf{Y}) = \int_\Omega \left( \frac{1}{2\varepsilon_0\varepsilon_r} |\mathbf{Y}|^2 + \frac{1}{2\mu_0} |\nabla \times \mathbf{A}|^2 \right) d\mathbf{r} \quad (4)$$

$$H^{qm}(\mathbf{A}, \psi, \psi^*) = \int_\Omega \left[ \psi^* \frac{(\mathbf{p}-e\mathbf{A})^2}{2m} \psi + \psi^* V \psi \right] d\mathbf{r} \quad (5)$$

Apparently, $H^{em}$ represents the Hamiltonian of the EM system, and $H^{qm}$ is the Hamiltonian of the QM system.

By employing the variational principle and decomposing the wave function ($\psi$) into the real and imaginary parts

$$\psi = \frac{1}{\sqrt{2\hbar}} (\psi_r + i\psi_i) \quad (6)$$

where the canonical Hamilton's equations for the EM-QM systems are deduced as

$$\frac{\partial \mathbf{A}}{\partial t} = \frac{\partial H}{\partial \mathbf{Y}}, \quad \frac{\partial \mathbf{Y}}{\partial t} = -\frac{\partial H}{\partial \mathbf{A}} \quad (7)$$

$$\frac{\partial \psi_r}{\partial t} = \frac{\partial H}{\partial \psi_i}, \quad \frac{\partial \psi_i}{\partial t} = -\frac{\partial H}{\partial \psi_r} \quad (8)$$

In this infinite-dimensional Hamiltonian system, the canonical pairs are ($\psi_r$, $\psi_i$) and ($\mathbf{A}$, $\mathbf{Y}$), and their canonical equations are

$$\frac{\partial \mathbf{A}}{\partial t} = \frac{\mathbf{Y}}{\varepsilon_0 \varepsilon_r} \quad (9)$$

$$\frac{\partial \mathbf{Y}}{\partial t} = -\frac{\nabla \times \nabla \times \mathbf{A}}{\mu_0} + \mathbf{J} \quad (10)$$

$$\frac{\partial \psi_r}{\partial t} = \frac{(x_1 - x_2)}{i\sqrt{2\hbar}} \psi_r + \left( \frac{(x_1 + x_2)}{\sqrt{2\hbar}} + \frac{\sqrt{2}V}{\sqrt{\hbar}} \right) \psi_i \quad (11)$$

$$\frac{\partial \psi_i}{\partial t} = -\left( \frac{(x_1 + x_2)}{\sqrt{2\hbar}} + \frac{\sqrt{2}V}{\sqrt{\hbar}} \right) \psi_r + \frac{(x_1 - x_2)}{i\sqrt{2\hbar}} \psi_i \quad (12)$$

with

$$x_1 = \frac{(\mathbf{p} - e\mathbf{A})^2}{2m}, x_2 = \frac{(\mathbf{p} + e\mathbf{A})^2}{2m}$$

and the quantum current $\mathbf{J}$ is described by the following form

$$\mathbf{J} = \frac{e}{2m} [\psi^*(\mathbf{p} - e\mathbf{A})\psi + \psi(-\mathbf{p} - e\mathbf{A})\psi^*] \quad (13)$$

The current term is generated by the motion of atoms and will radiate EM fields. The generated quantum current will be coupled back to the Maxwell's equations.

For QM part, equations (11) and (12) can be written as

$$\frac{\partial}{\partial t} \begin{pmatrix} \psi_r \\ \psi_i \end{pmatrix} = (\mathbf{L}) \begin{pmatrix} \psi_r \\ \psi_i \end{pmatrix} \quad (14)$$

according to the matrix $\mathbf{L}$, it is easy to get

$$\begin{aligned} \mathbf{L}^T &= -\mathbf{L} \\ \mathbf{L}^T \mathbf{E} + \mathbf{E}\mathbf{L} &= 0 \\ \mathbf{E} &= \begin{pmatrix} \mathbf{0} & \mathbf{I} \\ -\mathbf{I} & \mathbf{0} \end{pmatrix} \end{aligned} \quad (15)$$

Hence, equation (1) possesses symplectic structure, and the symplectic algorithm can be employed to calculate the QM part.

For the temporal direction, the EM part in homogenous, loss-less, and sourceless medium can be described by

$$\frac{\partial}{\partial t} \begin{pmatrix} \mathbf{A} \\ \mathbf{Y} \end{pmatrix} = \begin{pmatrix} \mathbf{0}_{3\times 3} & \boldsymbol{\varepsilon}_{3\times 3}^{-1} \\ -(\mu^{-1}\nabla \times \nabla \times)_{3\times 3} & \mathbf{0}_{3\times 3} \end{pmatrix} \begin{pmatrix} \mathbf{A} \\ \mathbf{Y} \end{pmatrix} = \mathbf{U} \begin{pmatrix} \mathbf{A} \\ \mathbf{Y} \end{pmatrix} \quad (16)$$

As the matrix $\mathbf{U}$ can be represented as

$$\mathbf{U} = \begin{pmatrix} \mathbf{0}_{3\times 3} & \mathbf{U}_A \\ \mathbf{U}_B & \mathbf{0}_{3\times 3} \end{pmatrix} \quad (17)$$

where

$$\mathbf{U}_A = \boldsymbol{\varepsilon}_{3\times 3}^{-1} = \begin{pmatrix} 1/\varepsilon_0\varepsilon_r & 0 & 0 \\ 0 & 1/\varepsilon_0\varepsilon_r & 0 \\ 0 & 0 & 1/\varepsilon_0\varepsilon_r \end{pmatrix} = \mathbf{U}_A^T \quad (18)$$

TABLE I
The coefficients $c_l=c_{m+1-l}(0<l<m+1); d_l=d_{m-l}(0<l<m), d_m=0$ in the symplectic algorithms.

| Coefficients | 2nd-order 2-stage | 4th-order 5-stage |
|---|---|---|
| $c_1$ | 0.5 | 0.17399689146541 |
| $d_1$ | 1.0 | 0.62337932451322 |
| $c_2$ | 0.5 | -0.12038504121430 |
| $d_2$ | 0.0 | -0.12337932451322 |
| $c_3$ | - | 0.89277629949778 |
| $d_3$ | - | $d_2$ |

TABLE II
The coefficients for different-order accurate collocated differences.

| Order() | $\beta_{-2}$ | $\beta_{-1}$ | $\beta_0$ | $\beta_1$ | $\beta_2$ |
|---|---|---|---|---|---|
| 2 | - | 1 | -2 | 1 | - |
| 4 | -1/12 | 4/3 | -5/2 | 4/3 | -1/12 |

$$\boldsymbol{U}_B = -(\mu^{-1}\nabla \times \nabla \times)_{3\times 3}$$
$$= \begin{pmatrix} \left(\frac{\partial^2}{\partial y^2}+\frac{\partial^2}{\partial z^2}\right) & -\frac{\partial^2}{\partial x \partial y} & -\frac{\partial^2}{\partial x \partial z} \\ -\frac{\partial^2}{\partial x \partial y} & \left(\frac{\partial^2}{\partial x^2}+\frac{\partial^2}{\partial z^2}\right) & -\frac{\partial^2}{\partial y \partial z} \\ -\frac{\partial^2}{\partial x \partial z} & -\frac{\partial^2}{\partial y \partial z} & \left(\frac{\partial^2}{\partial x^2}+\frac{\partial^2}{\partial y^2}\right) \end{pmatrix}$$
$$= \boldsymbol{U}_B^T \quad (19)$$

According to (18) and (19), we have
$$\boldsymbol{U}^T\boldsymbol{J}_{6\times 6} + \boldsymbol{J}_{6\times 6}\boldsymbol{U}$$
$$= \begin{pmatrix} \boldsymbol{0}_{3\times 3} & \boldsymbol{U}_B^T \\ \boldsymbol{U}_A^T & \boldsymbol{0}_{3\times 3} \end{pmatrix}\begin{pmatrix} \boldsymbol{0}_{3\times 3} & \boldsymbol{I}_{3\times 3} \\ -\boldsymbol{I}_{3\times 3} & \boldsymbol{0}_{3\times 3} \end{pmatrix} + \begin{pmatrix} \boldsymbol{0}_{3\times 3} & \boldsymbol{I}_{3\times 3} \\ -\boldsymbol{I}_{3\times 3} & \boldsymbol{0}_{3\times 3} \end{pmatrix}\begin{pmatrix} \boldsymbol{0}_{3\times 3} & \boldsymbol{U}_A \\ \boldsymbol{U}_B & \boldsymbol{0}_{3\times 3} \end{pmatrix}$$
$$= \begin{pmatrix} -\boldsymbol{U}_B^T + \boldsymbol{U}_B & \boldsymbol{0}_{3\times 3} \\ \boldsymbol{0}_{3\times 3} & \boldsymbol{U}_A^T - \boldsymbol{U}_A \end{pmatrix} = \begin{pmatrix} \boldsymbol{0}_{3\times 3} & \boldsymbol{0}_{3\times 3} \\ \boldsymbol{0}_{3\times 3} & \boldsymbol{0}_{3\times 3} \end{pmatrix} \quad (20)$$

this means that $\boldsymbol{U}$ is an infinitesimal real symplectic matrix.

Additionally,
$$\boldsymbol{U} = \begin{pmatrix} \boldsymbol{0}_{3\times 3} & \boldsymbol{U}_A \\ \boldsymbol{0}_{3\times 3} & \boldsymbol{0}_{3\times 3} \end{pmatrix} + \begin{pmatrix} \boldsymbol{0}_{3\times 3} & \boldsymbol{0}_{3\times 3} \\ \boldsymbol{U}_B & \boldsymbol{0}_{3\times 3} \end{pmatrix} = \boldsymbol{U}_1 + \boldsymbol{U}_2 \quad (21)$$

since $\boldsymbol{U}_1^\varsigma = 0, \boldsymbol{U}_2^\varsigma = 0, \varsigma \geq 2$, then
$$\exp(\Delta t \boldsymbol{U}_1) = \boldsymbol{I}_{6\times 6} + \Delta t \boldsymbol{U}_1 \quad (22)$$
$$\exp(\Delta t \boldsymbol{U}_2) = \boldsymbol{I}_{6\times 6} + \Delta t \boldsymbol{U}_2 \quad (23)$$

therefore, $\exp(\Delta t \boldsymbol{U}_1)$ and $\exp(\Delta t \boldsymbol{U}_2)$ can be obtained explicitly. Note that the matrices $\boldsymbol{U}_1$ and $\boldsymbol{U}_2$ do not commute
$$\boldsymbol{U}_1\boldsymbol{U}_2 \neq \boldsymbol{U}_2\boldsymbol{U}_1 \quad (24)$$

Accordingly, the symplectic integrator is applicable to solve the EM subsystem [27].

Obviously, the symplectic algorithm can be constructed to solve the above M-S system. Hence, utilizing symplectic mapping, the solution of equation (9), (10), (11) and (12) for the M-S system from $t = 0$ to $t = \Delta t$ is established approximately
$$\exp(\Delta t \boldsymbol{F}) = \prod_{l=1}^{m} \exp(c_l \Delta t \boldsymbol{F}_1) \exp(d_l \Delta t \boldsymbol{F}_2) + O(\Delta t^{p+1})$$
$$= \prod_{l=1}^{m} (1 + c_l \Delta t \boldsymbol{F}_1)(1 + d_l \Delta t \boldsymbol{F}_2) + O(\Delta t^{p+1}) \quad (25)$$

where $\boldsymbol{F} = \boldsymbol{L}$ or $\boldsymbol{U}$ and $c_l$ and $d_l$ are the coefficients of symplectic integration algorithm. $m$ is the number of the sub-step in each time step and $p$ represents the approximation order, generally $m \geq p$. In order to achieve the value of $p$ as high as possible for a given number $m$, some efforts have been made to obtain the coefficients $c_l$ and $d_l$. Substantial numerical experiments prove that the coefficients $c_l$ and $d_l$ in [27] perform better than other symplectic integrators. The coefficients for different order approximations are listed in Table I.

### B. High-Order Symplectic Algorithm for EM System

In section $A$, we have introduced that the EM and QM systems can be calculated by the symplectic algorithm, hence the self-consistent solution for the M-S system ((11), (12) and the current term **J**) can be solved by using $m$-stage and fourth-order symplectic integrators. Therefore, the components of M-S equations at the discrete points in time and space can be described as
$$G(i,j,k,t) = G^{n+l/m}(i\Delta x, j\Delta y, k\Delta z, (n+l/m)\Delta t) \quad (26)$$
where $\Delta x$, $\Delta y$, $\Delta z$ are the cell sizes along three different directions.

In addition, some studies have indicated that the satisfactory results can be obtained by using a combination of the high-order time algorithm and the high-order space algorithm [28]. Based on this, the explicit fourth-order collocated difference is adopted for discretizing the differential operators in space domain.

From above descriptions, the $q$th-order accurate collocated difference expressions, which are applied to discretize the second-order differential operators, are of the following forms
$$\frac{\partial^2 G^{n+l/m}(r)}{\partial \zeta^2} = \frac{1}{\Delta_\zeta^2} \sum_{d=-\frac{q}{2}}^{d=\frac{q}{2}} \beta_d G^{n+l/m}(r+d) + O(\Delta_\zeta^{q+1}) \quad (27)$$

where $\zeta = x, y, z$, $r = i, j, k$, and $\beta_d$ listed in Table II denotes the difference coefficients at different grid points for the second and fourth-order differences.

Based on the above descriptions, the SFDTD(4,4) update equations can be given by (taking the $y$-direction as an example).

$$A_y^{n+\frac{l}{m}}\left(i, j+\frac{1}{2}, k\right)$$
$$= A_y^{n+\frac{l-1}{m}}\left(i, j+\frac{1}{2}, k\right) + \frac{c_l \Delta t}{\varepsilon_0 \varepsilon_r} Y_y^{n+\frac{l-1}{m}}\left(i, j+\frac{1}{2}, k\right) \quad (28)$$

$$Y_y^{n+\frac{l}{m}}\left(i, j+\frac{1}{2}, k\right)$$
$$= Y_y^{n+\frac{l-1}{m}}\left(i, j+\frac{1}{2}, k\right) + J_y^{n+\frac{l-1}{m}}\left(i, j+\frac{1}{2}, k\right)$$
$$+ \alpha_{x1} \times \left[A_y^{n+\frac{l-1}{m}}\left(i+1, j+\frac{1}{2}, k\right) - 2A_y^{n+\frac{l-1}{m}}\left(i, j+\frac{1}{2}, k\right)\right.$$
$$\left. + A_y^{n+\frac{l-1}{m}}\left(i-1, j+\frac{1}{2}, k\right)\right]$$
$$+ \alpha_{x2} \times \left[A_y^{n+\frac{l-1}{m}}\left(i+2, j+\frac{1}{2}, k\right) - 2A_y^{n+\frac{l-1}{m}}\left(i, j+\frac{1}{2}, k\right)\right.$$
$$\left. + A_y^{n+\frac{l-1}{m}}\left(i-2, j+\frac{1}{2}, k\right)\right]$$
$$+ \alpha_{z1} \times \left[A_y^{n+\frac{l-1}{m}}\left(i, j+\frac{1}{2}, k+1\right) - 2A_y^{n+\frac{l-1}{m}}\left(i, j+\frac{1}{2}, k\right)\right.$$

$$+ A_y^{n+\frac{l-1}{m}}\left(i, j+\frac{1}{2}, k-1\right)\Big]$$
$$+ \alpha_{z2} \times \left[A_y^{n+\frac{l-1}{m}}\left(i, j+\frac{1}{2}, k+2\right) - 2A_y^{n+\frac{l-1}{m}}\left(i, j+\frac{1}{2}, k\right)\right.$$
$$\left.+ A_y^{n+\frac{l-1}{m}}\left(i, j+\frac{1}{2}, k-2\right)\right] \quad (29)$$

where $\alpha_{x1} = \frac{4}{3} d_l \times \kappa_x$, $\alpha_{z1} = \frac{4}{3} d_l \times \kappa_z$, $\alpha_{x2} = -\frac{1}{12} d_l \times \kappa_x$, $\alpha_{z2} = -\frac{1}{12} d_l \times \kappa_z$. For the cubic grid, we have $\Delta x = \Delta y = \Delta z = \Delta$ and $\kappa_x = \kappa_y = \kappa_z = \kappa_\zeta$. The constant $\kappa_\zeta$ is defined as a kind of the Courant-Friedrichs-Levy (CFL) coefficients in the FDTD approach.

### C. High-Order Symplectic Algorithm for QM System

In theory, (9-13) compose a complete symplectic framework, where four variables $\mathbf{A}$, $\mathbf{Y}$, $\psi_r$, $\psi_i$ are calculated subsequently in each time step. In [18], to eliminate the computational burden of the numerical solver caused by the multiscale problem between QM and EM systems, a REE approach is presented. The wave function $\psi(\mathbf{r}, t)$ can be expressed as

$$\psi(\mathbf{r}, t) = C_g(t) e^{-iE_g t/\hbar} \psi_g(\mathbf{r}) + C_e(t) e^{-iE_e t/\hbar} \psi_e(\mathbf{r}) \quad (30)$$

where $C_g(t)$ and $C_e(t)$ denote the coefficients which satisfy the energy-conserving condition

$$|C_g(t)|^2 + |C_e(t)|^2 = 1 \quad (31)$$

the exponential terms of $e^{-iE_g t/\hbar}$ and $e^{-iE_e t/\hbar}$ characterize the time evolution of the eigenstates, and $E_g = \hbar \omega_g$, $E_e = \hbar \omega_e$. These two dominant atomic states are denoted by $\psi_g(\mathbf{r})$ (for ground) and $\psi_e(\mathbf{r})$ (for excited), respectively. The QM system is nothing but a two-level system, and the two states can be expressed by

$$\psi_g(\mathbf{r}_{s=x,y,x}) = \left(\frac{1}{a\sqrt{\pi}}\right)^{3/2} e^{-(x^2+y^2+z^2)/2a^2} \quad (32)$$

$$\psi_e(\mathbf{r}_{s=x,y,x}) = \left(\frac{1}{a\sqrt{\pi}}\right)^{3/2} \left(\frac{s}{a}\right) \sqrt{2} e^{-(x^2+y^2+z^2)/2a^2} \quad (33)$$

and $a = \sqrt{\hbar/m\omega}$.

At last, two sets of equations about $C_g(t)$ and $C_e(t)$ are deduced

$$i\hbar \frac{dC_g(t)}{dt} = -\frac{e\mathbf{A}}{m} \langle \psi_g | \mathbf{p} | \psi_e \rangle C_e(t) e^{-i\omega_0 t} + \frac{e^2 \mathbf{A}^2}{2m} C_g(t) \quad (34)$$

$$i\hbar \frac{dC_e(t)}{dt} = -\frac{e\mathbf{A}}{m} \langle \psi_e | \mathbf{p} | \psi_g \rangle C_g(t) e^{i\omega_0 t} + \frac{e^2 \mathbf{A}^2}{2m} C_e(t) \quad (35)$$

and the current term $\mathbf{J}$ is given below

$$\langle \mathbf{J} \rangle = -\frac{e^2 \mathbf{A}}{m} \left(|C_g(t)|^2 + |C_e(t)|^2\right)$$
$$+ \frac{e}{m} \left[C_g^*(t) C_e(t) e^{-i\omega_0 t} \langle \psi_g | \mathbf{p} | \psi_e \rangle + C_e^*(t) C_g(t) e^{i\omega_0 t} \langle \psi_e | \mathbf{p} | \psi_g \rangle\right] \quad (36)$$

where $\omega_0$ is the transition frequency which is equal to $\omega_e - \omega_g$.

The SFDTD discretization of (34), (35) and (36) are

$$C_g^{n+\frac{l}{m}} = C_g^{n+\frac{l-1}{m}} - \frac{eA_y \Delta t}{i\hbar m} \langle \psi_g | \mathbf{p} | \psi_e \rangle C_e^{n+\frac{l-1}{m}} e^{-i\omega_0 \left(n+\frac{l-1}{m}\right)\Delta t} + \frac{e^2 A_y^2 \Delta t}{2i\hbar m} C_g^{n+\frac{l-1}{m}} \quad (37)$$

$$C_e^{n+\frac{l}{m}} = C_e^{n+\frac{l-1}{m}} -$$

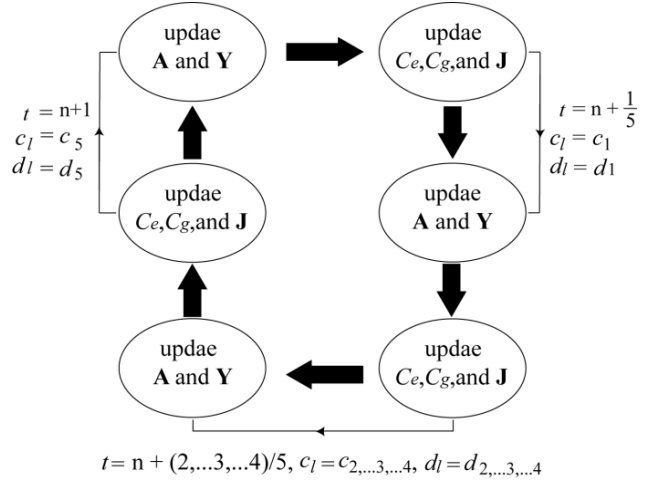

Fig. 1. The simulation procedures of the symplectic algorithm for the coupled M-S system.

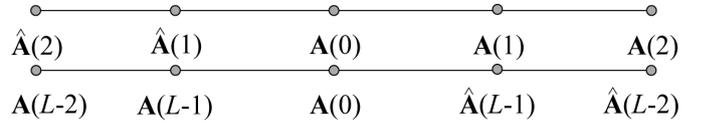

Fig. 2. Boundary treatment of $\mathbf{A}$ field by the image theory.

$$\frac{eA_y \Delta t}{i\hbar m} \langle \psi_e | \mathbf{p} | \psi_g \rangle C_g^{n+\frac{l-1}{m}} e^{i\omega_0 \left(n+\frac{l-1}{m}\right)\Delta t} + \frac{e^2 A_y^2 \Delta t}{2i\hbar m} C_e^{n+\frac{l-1}{m}} \quad (38)$$

$$J_y^{n+\frac{l}{m}} = -\frac{e^2 A_y}{m} \left(\left|C_g^{n+\frac{l-1}{m}}\right|^2 + \left|C_e^{n+\frac{l-1}{m}}\right|^2\right) +$$
$$\frac{e}{m}\left[\left(C_g^{n+\frac{l-1}{m}}\right)^* C_e^{n+\frac{l-1}{m}} e^{-i\omega_0\left(n+\frac{l-1}{m}\right)\Delta t} \langle \psi_g | \mathbf{p} | \psi_e \rangle + \right.$$
$$\left. \left(C_e^{n+\frac{l-1}{m}}\right)^* C_g^{n+\frac{l-1}{m}} e^{i\omega_0\left(n+\frac{l-1}{m}\right)\Delta t} \langle \psi_e | \mathbf{p} | \psi_g \rangle\right] \quad (39)$$

Finally, equations (28), (29), (37), (38) and (39) constitute the high-order discretized equations for solving the coupled M-S system. Fig. 1 shows the simulation procedures of the symplectic algorithm for the coupled M-S system.

Although the canonical symplectic algorithm [23] which is proposed by Chen *et al.* is more suitable to solve the MS equations when simulating the problems of the high harmonic generation (HHG) physics and the stabilization effect of the ionization, our proposed is optimal for the oscillation problem. As the variables in QM part are coordinate independent, less memory and CPU time will be cost when using our method to solve the coupled MS system.

### D. Boundary Condition in EM system

To obtain the reversible population inversion of the atomic system, the Dirichlet boundary condition is indispensable. In the proposed M-S system, only $\mathbf{A}$ and $\mathbf{Y}$ components should be discretized in space, hence the Dirichlet boundary condition is only applied to the EM system. For simplicity, taking one dimension as an example, $\mathbf{A}(0) = \mathbf{A}(L) = 0$ and $\mathbf{Y}(0) = \mathbf{Y}(L) = 0$ are set at the two ends of the $\mathbf{A}$ and $\mathbf{Y}$ boundaries, respectively. $L$ denotes the resonant cavity size.

For the fourth-order collocated differences in space, the image theory applied to the EM fields (**E** and **H**) can be directly adopted to **A** and **Y** components. In this work, the image theory is used, for **A** and **Y** components, we have **A**(0) = 0, **Â**(1) = −**A**(1), and **Â** (2) = −**A**(2). **Â** (1) and **Â** (2) are the image points of **A**(1) and **A**(2), respectively (see Fig. 2). For another side, **A**(L) = 0, **Â** (L-1) = −**A**(L-1), and **Â** (L-2) = − **A**(L-2). **Â** (L-1) and **Â** (L-2) are the image points of **A**(L-1) and **A**(L-2), respectively. Similarly, for **Y** components, **Y**(0) = 0, **Ŷ** (1) = **Y**(1), and **Ŷ** (2) = **Y**(2); **Y**(L) = 0, **Ŷ** (L-1) = **Y**(L-1), and **Ŷ**(L-2) = **Y**(L-2).

## III. NUMERICAL RESULTS

To investigate the accuracy and efficiency of the proposed fourth-order SFDTD(4,4) algorithm for QM System, it will be employed to solve the coupled M-S system. The population inversion defined by the formula (40) is used to describe the Rabi oscillation phenomenon.

$$W(t) = |C_e(t)|^2 - |C_g(t)|^2 \quad (40)$$

An artificial atom is illuminated by the external fields in a metal resonating cavity without any dielectric and radiation loss. The size of the nanocavity is $L_x = L_y = L_z = 40$ nm and the grid size $\Delta x = \Delta y = \Delta z = 1$ nm. The resonating cavity is excited at its fundamental mode (TE$_{101}$), with the initial solution

$$Y_y|_{t=0} = -\varepsilon_r \varepsilon_0 E_0 \sin\left(\frac{\pi}{L_x}x\right)\sin\left(\frac{\pi}{L_z}z\right)\cos(\omega t)|_{t=0} \quad (41)$$

where

$$\omega = \sqrt{\left(\frac{\pi}{L_x}\right)^2 + \left(\frac{\pi}{L_x}\right)^2} \quad (42)$$

and the relative permittivity $\varepsilon_r$ equals to 1 in the vacuum. The atom is placed at the center of the structure and is in a superposed state with $C_g = 1/\sqrt{2}$ and $C_e = 1/\sqrt{2}$. Under the external illumination of EM fields, the population is transferred back and forth between the ground state and excited state cyclically. Most importantly, the atomic population evolution can be analytically described by the Rabi model [18], which provides a platform for verifying the accuracy of the proposed algorithm and lays a solid foundation for more complex applications by using the proposed symplectic algorithm.

### A. Effect of EM Field strength

Firstly, the case of exact resonance (Δ = 0) is considered, where Δ = ω − ω$_0$ represents the detuning between the atomic transition frequency ( ω$_0$ ) and the fundamental resonance frequency ( ω) of the cavity. In this situation, the relation between the Rabi frequency Ω$_R$ and the intensity of **E** field is linear (see Equation. (A.9) - (A.10) in [18]). Two cases with weak field ( Ω = 0.02 ω) and strong field ( Ω = 0.2 ω) excitations are considered, the time step is 6.75 × 10$^{-7}$ ns.. According to (32), the population inversion is computed and shown in Fig. 3(a) and Fig. 4(a) for the two cases. The population inversion obtained by the FDTD(2,2) approach and SFDTD(4,4) algorithm have good agreements with the Rabi model [18].

Additionally, for a comparative study between the SFDTD

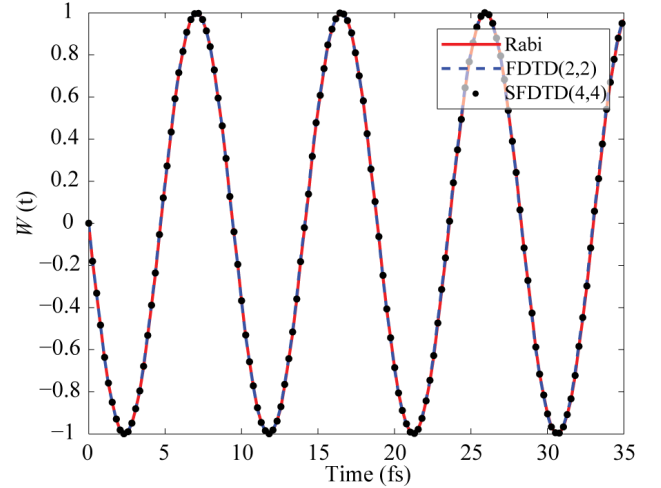

Fig. 3. (a) The calculation results for a weak field (Ω = 0.02ω) in the exact resonance condition.

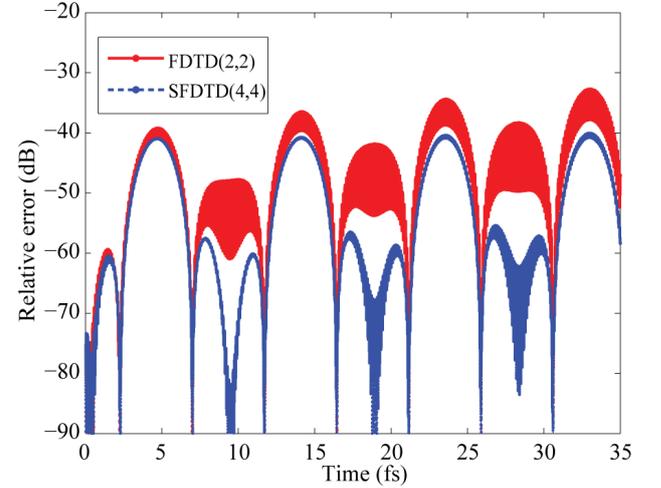

Fig. 3. (b) The relative errors of the FDTD(2,2) approach and SFDTD(4,4) algorithm for a weak field (Ω = 0.02ω) in the exact resonance condition.

(4,4) algorithm and the FDTD(2,2) approach, relative errors ($W_{rr}$) in dB for the population inversion are estimated by using the following formula

$$W_{rr} = 20\log_{10}\frac{|W(t) - W_{ref}(t)|}{max|W_{ref}(t)|} \quad (43)$$

where $W(t)$ refers to the numerical result and $W_{ref}(t)$ is the reference analytical solution. The denominator of (43) ($max|W_{ref}(t)|$) denotes the maximum value of the reference solution. For the sake of comparisons, the simulation results of the Rabi model are used as reference solutions. Relative errors of the proposed SFDTD(4,4) algorithm depicted in Fig. 3 (b) and Fig.4 (b) are reduced significantly compared to the traditional FDTD(2,2) approach in the whole simulation process. Furthermore, the relative error of the traditional FDTD(2,2) approach increases as the simulation time increases. However, in both cases, the relative error of the SFDTD(4,4) algorithm remains stable owing to its high-order precision. The results confirm that the SFDTD(4,4) algorithm has high accuracy and stability in solving the M-S system.

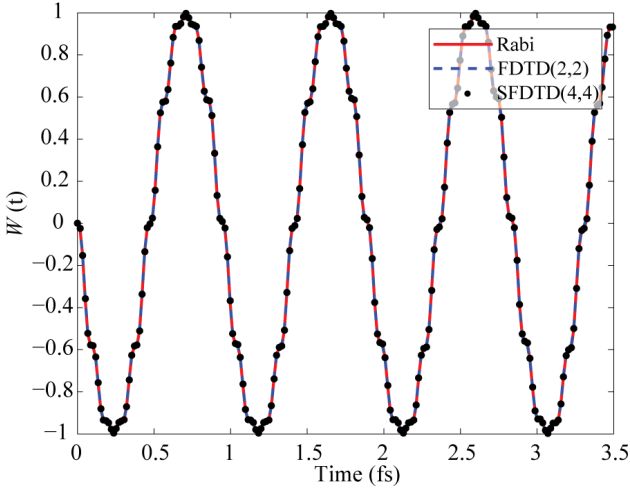

Fig. 4. (a) The calculation results for a strong field ($\Omega = 0.2\omega$) in the exact resonance condition, where rotating wave approximation (RWA) breaks down [29.]

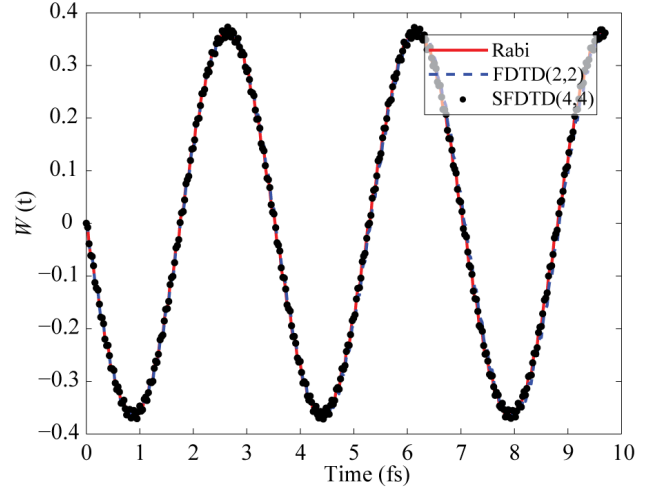

Fig. 5. (a) The calculation results for a small detuning $\Delta=0.05\omega$ condition.

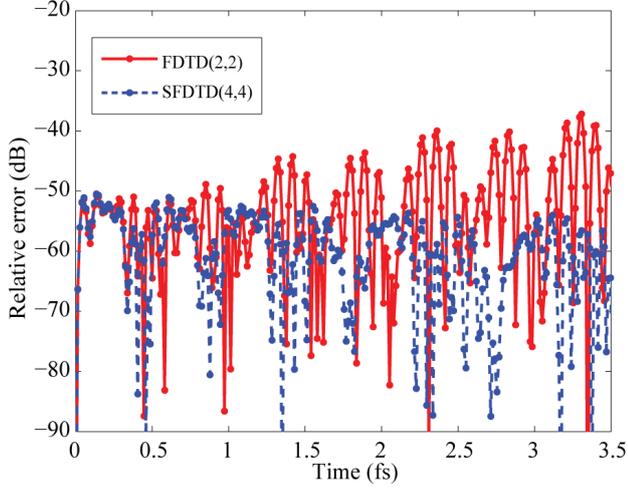

Fig. 4. (b) The relative errors of the FDTD(2,2) approach and SFDTD(4,4) algorithm for a strong field ($\Omega = 0.02\omega$) in the exact resonance condition, where rotating wave approximation (RWA) breaks down [29].

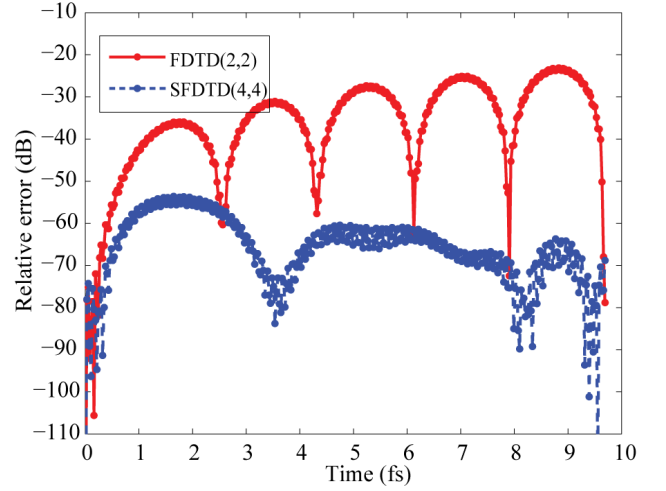

Fig. 5. (b) The relative errors of the FDTD(2,2) approach and SFDTD(4,4) algorithm for a small detuning $\Delta=0.05\omega$ condition.

### B. Effect of Detuning

Next, the effect of detuning factor $\Delta$ is considered. We set $\Omega = 0.02\omega$ and consider the following conditions where $\Delta = 0.05\omega$ and $\Delta = 0.3\omega$. Fig. 5 (a) and Fig. 6 (a) depict the population inversion for $\Delta = 0.05\omega$ and $\Delta = 0.3\omega$, respectively. It can be seen that the population inversion obtained by the FDTD(2,2) approach and the SFDTD(4,4) algorithm have good agreements with the analytical method for the detuning cases. Similarly, the relative errors of the two numerical approaches for the two cases ($\Delta = 0.05\omega$ and $\Delta = 0.3\omega$) are calculated and shown in Fig. 5 (b) and Fig. 6 (b), respectively. One can see that the proposed SFDTD(4,4) algorithm has smaller calculation error than the FDTD(2,2) approach.

It is worthy of noting that the analytical method in [18] must be modified for the detuning case (the detailed explanation appears in Appendix for reference).

Additionally, the grid size $\Delta = 0.5$ nm is also used in the low-order FDTD(2,2) approach. Accordingly, the time increment is set to be the half of the previous one. The relative errors for the FDTD(2,2) with dense grid (DG) ($\Delta = 0.5$ nm), FDTD(2, 2) with coarse grid ($\Delta =1$ nm) and the SFDTD(4, 4) with coarse grid ($\Delta = 1$ nm) are depicted in Fig. 7. It indicates that the accuracy of the FDTD(2,2) approach can be improved by increasing the grid resolution, but it is still worse than the SFDTD(4,4) algorithm. Furthermore, in terms of the FDTD(2,2) approach with dense grid, the CPU time is around 1203.716 s. However, the execution time of the SFDTD(4,4) algorithm is only 307.672 s. Simultaneously, the relative errors of the second-order FDTD(2,2) with different grid sizes always increase with the time evolution goes on. Hence, for the second-order FDTD(2,2) approach, the high-resolution settings in time and space can only slow down the increasing rate of the accumulated errors, but cannot eliminate the error accumulation.

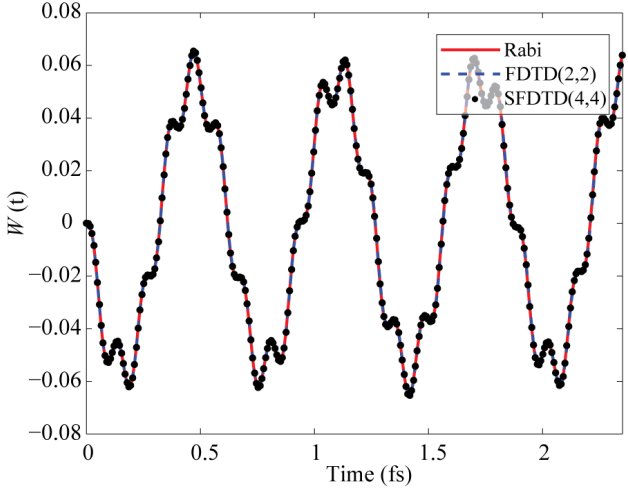

Fig. 6. (a) The calculation results for a large detuning Δ=0.3ω condition.

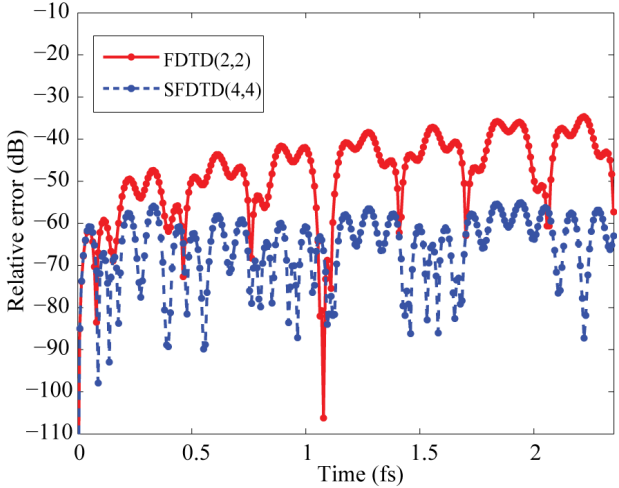

Fig. 6. (b) The relative errors of the FDTD(2,2) approach and SFDTD(4,4) algorithm for a large detuning Δ=0.3ω condition.

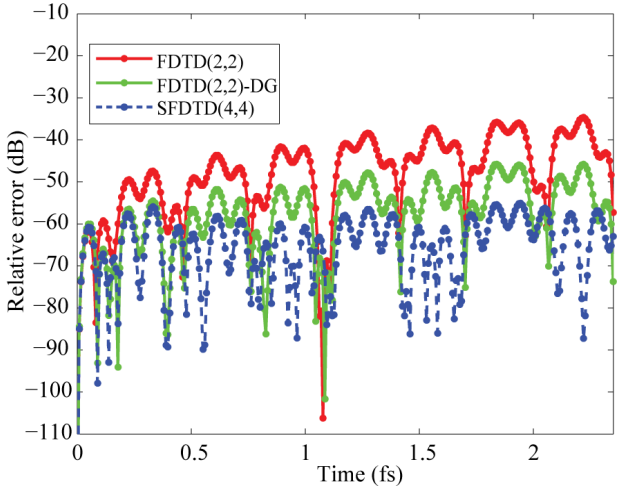

Fig. 7. The relative errors of the FDTD(2,2), FDTD(2,2)-DG and SFDTD(4, 4) approaches for a large detuning Δ=0.3ω condition. The FDTD(2,2) and SFDTD(4,4) use coarse grids of 1 nm. The FDTD(2,2)-DG uses fine grids of 0.5 nm.

## IV. CONCLUSION

A canonical symplectic structure of the M-S equations is deduced from the Hamiltonian of the EM-QM hybrid system. A three-dimensional fourth-order SFDTD(4,4) algorithm is proposed to solve the self-consistent M-S system which is essential to describe the EM field-artificial atom interaction. This method possesses fourth-order accuracy in space by utilizing the high-order collocated spatial differences and the image theory technique. Simultaneously, by applying symplectic integrators, the proposed algorithm exhibits superior numerical property when it was applied to compute the population inversion of an artificial atom in a resonant cavity. At last, a theoretical Rabi model which corresponds to the detuning case has been developed and provided to test the correctness of the presented high order numerical solver. The numerical examples validate the algorithm and demonstrate its long-term accuracy, stability, and efficiency. In future, this proposed high-order numerical solver will be applied to the electromagnetic field-artificial atom interaction in more complex environments where lossy media and irregular structures are included.

## APPENDIX

The Rabi model and the difference of $\mathbf{r} \cdot \mathbf{E}$ Hamiltonian and $\mathbf{p} \cdot \mathbf{A}$ Hamiltonian.

Under dipole approximation and "length" gauge, an electron interacts the external EM field can be described as

$$i\hbar \frac{\partial \psi(\mathbf{r}, t)}{\partial t} = \left\{ \frac{\mathbf{p}^2}{2m} + V(\mathbf{r}) - e\mathbf{r} \cdot \mathbf{E}(\mathbf{r}, t) \right\} \psi(\mathbf{r}, t) \quad (A.1)$$

where $e$ and $m$ denote the charge and the mass of the electron, respectively. The external EM field is assumed as a single mode field ($\mathbf{E} = \mathbf{E}_0 \cos(\omega t)$) which is treated classically and atom is modeled as a two-level quantum system [30]. Adopting the reduced eigenmode expansion technique, the coupled set of equations for the amplitudes $C_g(t)$ and $C_e(t)$ are derived as

$$i\hbar \frac{dC_g(t)}{dt} = -e\mathbf{E}_0 \cdot \langle \psi_g | \mathbf{r} | \psi_e \rangle C_e(t) \cos(t) e^{-i\omega_0 t} \quad (A.2)$$

$$i\hbar \frac{dC_e(t)}{dt} = -e\mathbf{E}_0 \cdot \langle \psi_e | \mathbf{r} | \psi_g \rangle C_g(t) \cos(t) e^{i\omega_0 t} \quad (A.3)$$

this is the theoretical Rabi model applied in [18].

Notice that the Hamiltonian of equation (A.1) can be expressed as

$$H = H_0 + H_1 \quad (A.4)$$

with

$$H_0 = \frac{\mathbf{p}^2}{2m} + V(\mathbf{r}) \quad (A.5)$$

$$H_1 = -e\mathbf{r} \cdot \mathbf{E}(\mathbf{r}, t) \quad (A.6)$$

this coupling Hamiltonian can be defined as $\mathbf{r} \cdot \mathbf{E}$ Hamiltonian.

However, according to equation (2) in this work, the atom-field Hamiltonian is expressed by the canonical momentum $\mathbf{p}$ and the vector potential $\mathbf{A}$ instead of the expression (A.4). According to [30], equation (2) corresponds to a Hamiltonian

$$H = H_0 + H_2 \quad (A.7)$$

with

$$H_2 = -\frac{e}{m}\mathbf{p} \cdot \mathbf{A}(\mathbf{r}, t) \quad (A.8)$$

and $H_0$ is given in (A.5). Additionally, the $\mathbf{A}^2$ term which is included in (A.7) is ignored as its value is usually small when compared to other terms.

According to the above analysis, the numerical solution system in this work is based on the $\mathbf{p} \cdot \mathbf{A}$ Hamiltonian. However, the theoretical Rabi model corresponds to $\mathbf{r} \cdot \mathbf{E}$ Hamiltonian. The relation between these two Hamiltonians has been proved to be related to the ratio of the atomic transition frequency ($\omega_0$) and the EM field frequency ($\omega$) (the detailed derivation is given in [30] page 150-151). For the case of exact resonance ($\Delta = 0$), the ratio R is equal to 1, there is no difference between these two Hamiltonians. On the contrary, for the detuning case, they are different, because the ratio R is no longer equal to 1. Hence, in order to be consistent with the $\mathbf{p} \cdot \mathbf{A}$ Hamiltonian, the update equations for $C_g(t)$ and $C_e(t)$ should be modified as

$$i\hbar\frac{dC_g(t)}{dt} = -e\mathbf{E}_0 \cdot \left(R \cdot \langle\psi_g|\mathbf{r}|\psi_e\rangle\right) C_e(t)\cos(t)e^{-i\omega_0 t} \quad (A.9)$$

$$i\hbar\frac{dC_e(t)}{dt} = -e\mathbf{E}_0 \cdot \left(R \cdot \langle\psi_e|\mathbf{r}|\psi_g\rangle\right) C_g(t)\cos(t)e^{i\omega_0 t} \quad (A.10)$$

The equations (A.9) and (A.10) provide a theoretical Rabi model which corresponds to the detuning case.